# Secure Data Access for Wireless Body Sensor Networks

Zhitao Guan[1], Tingting Yang[1], Xiaojiang Du[2], Mohsen Guizani[3]
1. School of Control and Computer Engineering, North China Electric Power University, China, guan@ncepu.edu.cn
2. Department of Computer and Information Science, Temple University, Philadelphia PA, USA, dxj@ieee.org
3. Qatar University, Doha, QATAR, mguizani@ieee.org

*Abstract*— Recently, with the support of mobile cloud computing, a large number of health-related data collected from various body sensor networks can be managed efficiently. However, to ensure data security and data privacy in cloud-integrated body sensor networks (C-BSN) is an important and challenging issue. In this paper, we present a novel secure access control mechanism MC-ABE (Mask Certificate–Attribute Based Encryption) for cloud-integrated body sensor networks. A specific signature is designed to mask the plaintext, then the masked data can be securely outsourced to cloud severs. An authorization certificate composing of the signature and related privilege items is constructed that is used to grant privileges to data receivers. To ensure security, a unique value is chosen to mask the certificate for each data receiver. The analysis shows that the proposed scheme has less computational cost and storage cost compared with other popular models.

## I. Introduction

Body sensor networks (BSNs) have emerged recently with the rapid development of wearable sensors, implantable sensors and short range wireless communication, which make pervasive healthcare monitoring and management become increasingly popular [1,2]. Using body sensor networks, health-related data of the patient can be collected and transferred to the healthcare staff in real time.

With the support of mobile cloud computing, cloud-integrated body sensor network (C-BSN) can be constructed [3]. In C-BSN, massive local body sensor networks are integrated together and mass data are collected and stored in cloud servers; healthcare issues will continually monitor their patients' status and exchange views when it is difficult to make diagnosis.

However, there are still several problems and challenges in C-BSN [3, 4]. For example, data security and data privacy must be a concern since patient-related data is private and sensitive. In this paper, we propose a secure data access control scheme named MC-ABE.

In this paper, we propose a novel secure access control mechanism MC-ABE to tackle with the aforementioned problems. The main contributions of this paper can be summarized as follows:

We construct one specific signature to CP-ABE to mask the plaintext, then realize secure encryption/decryption outsourcing. We construct the unique authentication certificate for each visitor, which makes the system achieve more effective control on malicious visitors. Our scheme takes less time than other compared methods to do data collecting, data transmission and data acquisition.

## II. Related Work

Recently, various techniques have been proposed to address the problems of data security and data privacy in C-BSN. In [5], Sahai and Waters proposed the Attribute-Based Encryption (ABE) to realize access control on encrypted data. In ABE, the ciphertext's encryption policy is associated with a set of attributes, and the data owner can be offline after data is encrypted. One year later, Goyal proposed a new type of ABE - key-policy attribute-based encryption (KP-ABE) [6]. In KP-ABE, the ciphertext's encryption policy is also associated with a set of attributes, but the attributes are organized into a tree structure (named access tree). The benefit of this approach is that a more flexible access control strategy can be attained and a fine-grained access control can be realized. Benthcourt proposed CP-ABE (ciphertext-policy attribute-based encryption) [7], in which the data owner constructed the access tree together with the visitors' identity information. The user can decrypt the ciphertext if and only if attributes in his private key match the access tree. Yu *et al*. [8] proposed the scheme based on KP-ABE, and combines with the two ore-encryption. It was proved that the proposed scheme can meet the security requirement in cloud quite well. Similarly, Wang *et al*. proposed an access control scheme based on CP-ABE, which is also secure and efficient in the cloud environment [9].

In [10], to reduce computation overhead and achieve secure encryption/decryption outsourcing, a portion of computation overhead was transferred from the data owner to the cloud sever. A similar method is also adopted in the work of Zhou [11], which proposed an efficient data management model to balance communication and storage overhead to reduce the cost of data management operations. In [12], Yao el al. proposed a novel access control mechanism in which data operation privileges are granted based on authorization certificates. The advantage of such mechanisms is that the computation cost can be decreased remarkably, since there is no bilinear map calculation. The disadvantage is that many of operations need to be handled by the data owner. In [13], the authors considered the problem of patient self-controlled access privilege to highly sensitive Personal Health Information. They proposed a Secure Patient-centric Access Control scheme which allows data requesters to have different access privileges based on their roles, and then assigns different attribute sets to them. However, they took the cloud server as trusted, and their scheme did not work well for user revocation.

## III. PRELIMINARIES

### A. Notations

In table 1, the notations used in MC-ABE are listed.

TABLE I. NATIONS IN MC-ABE

| Acronym | Descriptions |
|---|---|
| DO | Data Owner |
| DR | Data Requester/Receiver |
| ESP | Encryption Service Provider |
| DSP | Decryption Service Provider |
| SSP | Storage Service Provider |
| TA | Trust Authority |
| SetS | Setup Sever |
| PK | Public Key |
| MK | Master Key |
| SK | Secret Key |
| M | Plaintext |
| CT | Ciphertext |
| T | Access Tree |
| MM | Masked Plaintext |
| Cert | Authorization Certificate |
| MValue | Mask Value |
| MCert | Masked Cert |

### B. Basics

*1) Bilinear Pairing*

Let G1 and G2 be two multiplicative cyclic groups of prime order p. Let g be a generator of G1 and e be a bilinear map, e: G1 × G1 →G2. For $a, b \in Z_p$, the bilinear map e has the following properties [3]:

1. Bilinearity: for all $u, v \in G_1$ $u, v \in G_1$, we have $e(u^a, v^b) = e(u,v)^{ab}$.

2. Non-degeneracy: $e(g,g) \neq 1$.

3. Symmetric: $e(g^a, g^b) = e(g,g)^{ab} = e(g^b, g^a)$.

*2) Discrete Logarithm (DL) Problem*

**Definition 1**: Discrete Logarithm (DL) Problem

Let G be a multiplicative cyclic group of prime order p and g be its generator, for all $\alpha \in Z_p$, given $g, g^\alpha$ as input, output $\alpha$.

The DL assumption holds in G if it is computationally infeasible to solve the DL problem in G [14].

### C. Access Structure

The access structure in CP-ABE is the tree-structure, which is named as access tree [2]. For the access tree T, the leaf nodes are associated with descriptive attributes; each interior node is a relation function, such as AND (n of n), OR (1of n), n of m (m>n).

Each DR has a set of attributes, which are associated with DR's SK. If DR's attributes set satisfies the access tree, the encrypted data can be decrypted by DR's SK.

### D. Assumption

In this work, we make the following assumptions.

Assumption 1: service providers (ESP, DSP, SSP) are semi-trusted. That is, they will follow our proposed protocol in general, but try to find out as much secret information as possible.

Assumption 2: SetS and TA are trusted. On no conditions will they leak information about data and related keys.

In order to deduce more information about encrypted data, service providers might combine their information to perform collusion attack. In our scheme, collusions between service providers are taken into consideration.

## IV. MC-ABE

### A. Overview

Our proposed scheme MC-ABE is shown in figure 2. Seven algorithms are included in MC-ABE: Setup, Encrypt$_{DO}$, Encrypt$_{ESP}$, KeyGen, CerGen, Decrypt$_{DSP}$, Decrypt$_{DR}$.

For data outsourcing, DO encrypts M with algorithm Encrypt$_{DO}$, in which a signature is used to mask M. Then ESP encrypts T with the algorithm Encrypt$_{ESP}$ to finish the encryption. The encrypted data is stored in SSP.

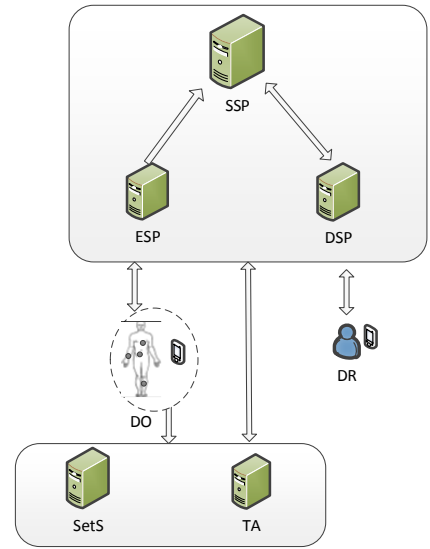

Fig. 1. System model

For data access, when DR requests data from SSP, the request is sent to TA after verification. TA chooses a unique value to the mask certificate for DR. Then, TA computes SK with the algorithm KeyGen. After that, SK is sent to DSP and the certificate is sent to DR. At the same time, SSP sends the CT to DSP. With SK and CT, DSP can do decryption and get M, which is masked by the signature. Once DR receives the certificate, one decrypts the masked certificate with his unique value (TA sends the unique value to this DR when the first authorized request occurs. It will be used in the following requests until this DR is revoked) to get the certificate. Using the certificate, DR can decrypt the masked M with signatures in the certificate.

In addition, if a DR is revoked, TA will mark the DR as 'revoked' and this DR's unique mask value will be invalid. No certificate will be granted to this DR any more.

### B. Two Important Notions

*1) Authorization Certificate (Cert)*

As shown in table 2, it includes five items that are privilege related information. DO provides the certificate related information to TA, then TA constructs the unique authorization certificate for each authorized DR.

TABLE II. STRUCTURE OF AUTHORIZATION CERTIFICATE

| File ID list (f1, f2…) |
| Valid Period(From the start time to the end time) |
| Signature ({signf1}, {signf2}…) |
| Privilege ({pf1}, {pf2}…) |
| PK, MK |

File ID: ID list of the authorized files.

Valid Period: it denotes the valid period of the signature, from the start time to the end time.

Signature: it is used by DO to mask the plaintext in data encryption; it is used by DR to get the plaintext in data decryption.

Privilege: the privilege denoted by the signature such as read, modify, delete.

PK, MK: These two keys are noted in table 1.

*2) Mask Value (MValue)*

The mask value is maintained by TA. For each DR, TA sets a unique mask value for him. The mask value is used to blind the authorization certificate before the certificate is sent to DR.

TABLE III. MASK VALUE TABLE (MAINTAINED BY TA)

| DRID | Mask value | Revocation |
|---|---|---|
| DR1 | $MValue_{DR1}$ | N |
| DR2 | $MValue_{DR2}$ | Y |
| DR3 | $MValue_{DR3}$ | N |

**DRID**: ID of DR.
**Mask value**: unique mask value for each DR.
**Revocation**: revocation mark. 'Y' means this DR is revoked. 'N' means this DR is authorized.

After TA receives a data access request, it checks DRID firstly. If the requester is a new user, TA generates a random number $t_{DRID} \in Z_p$ and inserts it into the mask value table. TA invokes the algorithm CerGen to compute the masked certificate.

Algorithm: CerGen($t_{DRID}$, PK)→MCert

Construct a certificate Cert as table 2 shows. MCert is the masked Cert.

Then, compute as follows:

$MValue = g^{t_{DRID}}$

$MCert = Cert \cdot e(g^\theta, g^{t_{DRID}}) = Cert \cdot e(g,g)^{\theta t_{DRID}}$

If DR is a new user, MValue and MCert will be sent to him. Otherwise, send MCert to the DR.

### C. Scheme Description

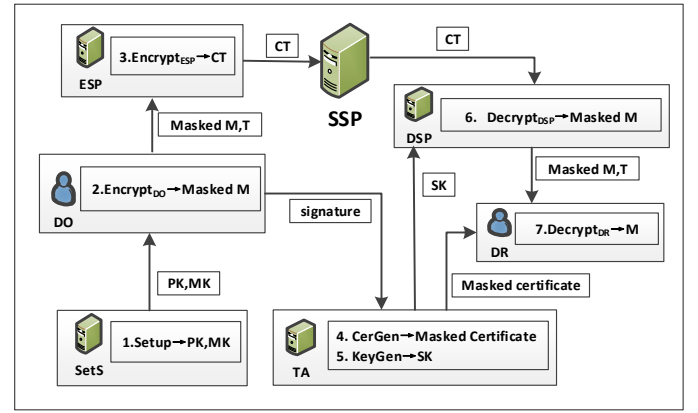

Fig. 2. Algorithms implementation in MC-ABE

The whole process of MC-ABE is shown in figure 3. In this section, we describe each step in detail.

*1) Data Outsourcing*
Firstly, Algorithm 1. Setup→ PK, MK

SetS performs the algorithm. Let G0 be a multiplicative cyclic group of prime order p and g be its generator, and four random numbers $\alpha, \beta, \varepsilon, \theta \in Z_p$ (Further details in [7]).

$PK = (G_0, g, h = g^\beta, e(g,g)^\alpha, g^\varepsilon, g^\theta)$

$MK = (\beta, g^\alpha)$

Secondly, Algorithm 2. $Encrypt_{DO}$(PK, M, K)→MM:

DO implements the algorithm.

For $k \subset K$ ( K is the set of operation privileges), we choose a random number $v_k \subset Z_p$, and then compute the signature:

$signature_k = e(g^\varepsilon, g^{v_k}) = e(g,g)^{\varepsilon v_k}$

For simplicity, let v denote the set of $v_k : v = \{v_k \mid k \in K\}$, signature denote the set of $signature_k$ :

$signature = \{signature_k \mid k \in K\}$ .

Choose a random number $s \in Z_p$, then

$$MM = \tilde{C} = M \cdot e(g,g)^{as} \, signature$$
$$= M \cdot e(g,g)^{as} \cdot e(g,g)^{\varepsilon v}$$

Lastly, Algorithm 3. EncryptESP(PK, s, T, MM) [7,11] → CT:

Implemented by ESP, the access tree T is encrypted from the root node R to leaf nodes. For each node x in T, choose a polynomial $q_x$, for node x,

$k_x$: the threshold value of x

$d_x$: the degree of $q_x$, $d_x = k_x - 1$

parent(x): a function returns the parent node of x.

$num_x$: number of child nodes of x. For a child node y, y is uniquely identified by an index number index(y), and $1 \leq index(y) \leq num_x$

$$q_x(0) = q_{parent(x)}(index(x))$$

For root node R, $q_R(0) = s$. Choose $d_R$ other points randomly to completely define $q_R$. For any other node x in T, let $q_x(0) = q_{paerent(x)}(index(x))$, and choose $d_x$ other points randomly to completely define $q_x$.

Y is the set of leaf nodes in T. Compute as follows:

$$C = h^s, \forall y \in Y : C_y = g^{q_y(0)}, C_y' = H(att(y))^{q_y(0)}$$

Then,

$$CT = \{T, \tilde{C} = M \cdot e(g,g)^{as} \cdot e(g,g)^{\varepsilon v}, C = h^s,$$
$$\forall y \in Y : C_y = g^{q_y(0)}, C_y' = H(att(y))^{q_y(0)}\}$$

CT is stored in SSP.

*2) Data Request*

When a DR requests data from SSP, TA generates SK and a certificate for DR. Most of the decryption cost is taken by DSP but DSP can not get M. Based on the effort of DSP, DR finishes the last step of decryption and gets M. Similar to data outsourcing, there are also three steps for data outsourcing.

First, TA generates SK for DR.

Algorithm 4. KeyGen(MK, S)→SK

S is the attributes set of DR. We generate a random number $r \in Z_p$, and then generate the random number $r_j \in Z_p$ for each $j \in S$. Compute as follows:

$$SK = (D = g^{(\alpha+r)/\beta},$$
$$\forall j \in S : D_j = g^r \cdot H(j)^{r_j}, D_j' = g^{r_j})$$

Then, TA sends SK to DSP.

Second, DSP performs the first step of data decryption: decrypt the access tree in CT to get MM.

Algorithm 5. DecryptDSP(SK, CT)→MM

When x is a leaf node, let i=att(x). Function att(x) denotes the attribute associated with the leaf node x in the tree.

If $i \in S$,

$$DecryptNodeL(CT, SK, x) = \frac{e(D_i, C_x)}{e(D_i', C_x')}$$

$$= \frac{e(g^r \cdot H(i)^{q_x(0)}, g^{q_x(0)})}{e(g^{r_i}, H(i)^{q_x(0)})} = e(g,g)^{r \cdot q_x(0)}$$

Otherwise,

$i \notin S$, $DecryptNodeL(CT, SK, x) = \perp$.

When x is an interior node, call the algorithm DecryptNodeNL(CT,SK,x).

For all of the children z of node x, call DecryptNodeL(CT, SK,z), and the output is $F_z$. Let $S_x$ be a $k_x$ (the threshold value of interior node) random set and let $F_z \neq \perp$. If no such set exists, the function cannot be satisfied, so return $\perp$.

Otherwise, compute as follows and return the result:

$$F_x = \prod_{z \in S_x} F_z^{\Delta_{i,S_x'}(0)}, where \begin{cases} i = index(z) \\ S_x' = \{index(z) : z \in S_x\} \end{cases}$$

$$= \prod_{z \in S_x} (e(g,g)^{r \cdot q_z(0)})^{\Delta_{i,S_x'}(0)}$$

$$= \prod_{z \in S_x} (e(g,g)^{r \cdot q_{parent(z)}(index(z))})^{\Delta_{i,S_x'}(0)}$$

$$= \prod_{z \in S_x} (e(g,g)^{r \cdot q_x(i)})^{\Delta_{i,S_x'}(0)}$$

$$= e(g,g)^{r \cdot q_x(0)}$$

Especially for root node R,

$$A = e(g,g)^{rq_T(0)} = e(g,g)^{r \cdot s}$$

Finally, $\tilde{C}_k \, / \, (e(C,D) \, / \, A)$

$$= \tilde{C} / (e(h^s, g^{(\alpha+r)/\beta}) / e(g,g)^{r \cdot s}) = M \cdot signature$$

Then, $M \cdot signature$ is sent to DR.

Receiving $M \cdot signature$ and MCert, DR implements the algorithm Decrypt$_{DR}$ to finish data decryption.

Lastly, DR removes the masked value in MM to get M.

Algorithm 6. Decrypt$_{DR}$($M \cdot signature$, MCert)→M

DR retrieves the Cert to get related signatures,

$$MCert / e(g^\theta, g^{t_{DRID}}) = Cert \cdot e(g,g)^{\theta t_{DRID}} / e(g,g)^{\theta t_{DRID}}$$
$$= Cert$$

Then, DR gets M with the signature,

$$M \cdot signature / signature = M$$

*3) User revocation*

An invalid DR is a DR who is thought to be malicious, or whose certificate is expired. The invalid DR should be revoked from the authorized access list. In MC-ABE, we can remove the MValue record in table 3 to revoke DR. First, TA modifies the revoked DR's 'Revocation' item from 'N' to 'Y' in Mask Value Table. Second, the current signature must be updated to a new one. After these two steps, the invalid DR is revoked. When he requests new data, he will be taken as a newcomer (the signature is updated, and he does not have the new one), and TA will refuse his request since he is marked as revoked. For valid DR, they will get the new signature and access the system as usual.

## V. SECURITY ANALYSIS

### A. Encryption and decryption outsource

In this paper, M is masked by DO before it is sent to ESP. DO and authorized DR can get M. ESP and DSP can get MM (Masked M), but they can not deduce M from MM.

<u>Theorem 1</u>: The security in encryption & decryption in MC-ABE is no weaker than that of CP-ABE.

<u>Proof</u>: In algorithm Encrypt$_{ESP}$, ESP encrypts the access tree T with the parameter s T and MM.

$\tilde{C} = M \cdot e(g,g)^{as} \cdot signature = M \cdot e(g,g)^{as} \cdot e(g,g)^{\varepsilon v}$ Using PK and s, ESP can get $e(g,g)^{as}$, what ESP gets is $M \cdot e(g,g)^{\varepsilon v}$.

The encrypted data in CP-ABE is $\tilde{C} = M \cdot e(g,g)^{as}$, both of $\alpha$ and s are random, let $z = \alpha \cdot s$, z is also random, then $\tilde{C} = Me(g,g)^z$ is equal to $Me(g,g)^{\varepsilon v_k}$. According to the security proof in [7]; the structure of $\tilde{C} = M \cdot e(g,g)^{as}$ is secure to prevent the adversary from deducing M. Thus, $Me(g,g)^{\varepsilon v_k}$ in our scheme is secure. That is to say, ESP can't deduce M with $Me(g,g)^{\varepsilon v_k}$, and encryption outsourcing is secure in MC-ABE.

For DSP, it can decrypt CT using SK, and get the masked $M = M \cdot signature$. The information DSP gets is the same as ESP. So, in MC-ABE, data decryption outsourcing is also secure since it is similar to data encryption outsourcing.

### B. Revocation

If a DR is revealed to be malicious, he will be revoked from the authorized user list. We update the signature encrypted in CT, after that, as shown in the following, the revoked DR can not get authorized data any more.

Revoked signature held by DR: $signature = e(g,g)^{\varepsilon v_k}$

Updated signature: $signature' = e(g,g)^{\varepsilon v'_k}$

Masked M'= $M \cdot signature' = Me(g,g)^{v'_k}$

Masked $M'/signature = Me(g,g)^{\varepsilon v'_k} / e(g,g)^{\varepsilon v_k}$

$\qquad = Me(g,g)^{\varepsilon(v'_k - v_k)}$

Thus, MC-ABE is secure in revocation.

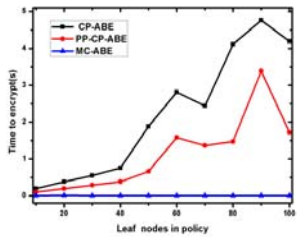

Fig.3(a). DO's computation cost for data encryption in CP-ABE, PP-CP-ABE and MC-ABE. In PP-CP-ABE, part of encryption computation is transferred to cloud sever to reduce DO's cost in MC-ABE.

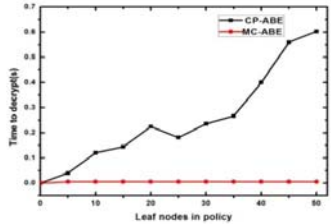

Fig.3(b). Computation cost of DO (The 95% confidence interval assuming random data with normal distribution is shown).

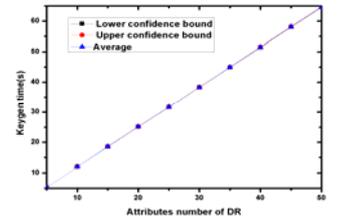

Fig.3(c). Computation cost of key generation (The 95% confidence interval assuming random data with normal distribution is shown).

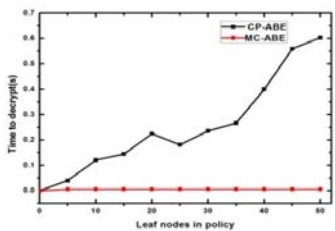

Fig.3(d). Computation cost of DR in CP-ABE and MC-ABE. Similar to ESP in MC-ABE, DSP also undertake most of the computation in decryption. The cost is proportional to attributes number in private key.

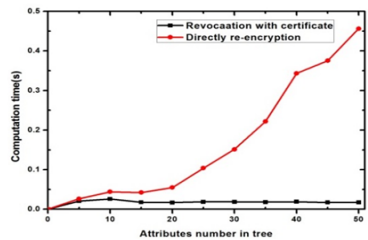

Fig.3(e). Computation cost for user revocation. With the authorization certificate in MC-ABE, revocation cost can be reduced obviously.

TABLE IV. CONFIDENCE INTERVAL OF KEY GENERATION COMPUTATION COST

| Att_num | CI | Ave |
|---|---|---|
| 10 | [11.909, 11.934] | 11.922 |
| 15 | [18.543, 18.589] | 18.566 |
| 20 | [25.127, 25.159] | 25.143 |
| 25 | [31.652, 31.731] | 31.691 |
| 30 | [38.265, 38.366] | 38.316 |
| 35 | [44.869, 44.956] | 44.912 |
| 40 | [51.455, 51.633] | 51.544 |
| 45 | [58.04, 58.1582] | 58.099 |
| 50 | [64.542, 64.678] | 64.610 |

The 95% confidence interval assuming random data with normal distribution is shown.

## VI. Performance Evaluation

### A. Numerical analysis

The main computation cost in the scheme is computations in algorithms, the attribute number in a tree or SK is the key factor to influence the computation cost. Simulation results of computation cost in MC-ABE are shown in figure 3. The confidence interval is shown in table 5 (Att_num indicates the number of DR's attributes. CI indicates the confidence interval, and Ave indicates the average value).

Compared to CP-ABE, more storage cost is incurred in MC-ABE because the certificate and the unique value are introduced. The items in certificates are related to data access privileges, so the storage space of the certificate is proportional to the number of the documents (data).

## VII. Conclusion

How to keep data security and data privacy in C-BSN is an important and challenging issue since the patients' health-related data is quite sensitive. In this paper, we propose a novel encryption outsourcing scheme MC-ABE that meets the requirements of data security and data privacy in C-BSN. In MC-ABE, one specific signature is constructed to mask the plaintext; the unique authentication certificate for each visitor is constructed; and, by performance evaluation, it shows that MC-ABE has less computation cost and storage cost compared with other popular models. In our future work, we plan to explore the possibility for improving the scalability of MC-ABE.

### Acknowledgment

This work is partially supported by Natural Science Foundation of China under grant 61402171, Central Government University Foundation under grant JB2014075, as well as US Army Research Office under grant WF911NF-14-1-0518.